\documentclass[prl,aps,showpacs,twocolumn]{revtex4}

\usepackage{epsfig}
\usepackage{latexsym}
\usepackage{graphicx}
\usepackage{amssymb}

\newcommand{\bi}{\begin{itemize}}
\newcommand{\ei}{\end{itemize}}
\newcommand{\be}{\begin{equation}}
\newcommand{\ee}{\end{equation}}

\newcommand{\ben}{\begin{eqnarray}}
\newcommand{\een}{\end{eqnarray}}
\newcommand{\benstar}{\begin{eqnarray*}}
\newcommand{\eenstar}{\end{eqnarray*}}

\begin{document}

\title{Local energy approach to the dynamic glass transition}

\author{Ivan Junier}
\affiliation{Departament de Fisica Fonamental, Facultat de Fisica, Universitat de Barcelona, Diagonal 647, 08028 Barcelona, Spain}

\begin{abstract}
We propose a new class of phenomenological models for dynamic glass transitions. The system consists of an ensemble
of mesoscopic regions to which {\it local} energies are allocated. 
At each time step, a region is randomly chosen and a new local energy   
is drawn from a distribution that {\it self-consistently} depends on the {\it global}
energy of the system. Then, the transition is accepted or not according to the Metropolis rule. 
Within this scheme, we model an energy {\it threshold} leading to a mode-coupling glass transition as in the $p$-spin
model.  The glassy dynamics is characterized by a two-step relaxation of the energy
autocorrelation function. 
The aging scaling is fully determined by the evolution of the global energy and 
linear violations of the fluctuation dissipation relation 
are found for observables
uncorrelated with the energies. Interestingly, 
our mean-field approach has a natural
extension to finite dimension, that we briefly discuss.
%Moreover, the scheme we propose can be adapted to glass forming liquids and
%hence may be a general approach to study the similarities and differences between the spin-glasses and structural glasses. 

\end{abstract}

\maketitle

Recent numerical studies \cite{scala,anglani,donati,grigera} have shed new light on the close connection between 
the dynamical slowing down of supercooled liquids 
and the topography of the underlying potential energy surface \cite{debe}.
In particular, a topological interpretation of the so-called mode-coupling temperature $T_M$ has been confirmed
\cite{anglani,donati,grigera}. 
That is, for temperatures $T$ smaller than $T_M$, the 
long-time relaxation is governed by activated processes between basins of potential energy minima. For
$T>T_M$, the thermal energy is larger than the potential energy barriers, which allows the system to freely explore 
its configuration space.

On the other hand, the (mean-field) mode-coupling approximation for liquids \cite{gotz} is exact for 
some disordered models \cite{bouchcu} such as the $p$-spin model \cite{cri0}.
The spherical version of the latter, endowed with a Langevin dynamics, gives a clear illustration of the interplay between 
the dynamics and the topography
of the energy landscape \cite{laloux}. The picture is the following \cite{laloux,cugkurch1}.
The stationary points whose energy is above the so-called {\it threshold} value $E_t$ are mainly
saddles, i.e. the minimum value of the Hessian eigenvalues, $\lambda_{m}$,  is negative. Those whose energy is below 
are mainly minima, i.e. $\lambda_{m}>0$, and at $E_t$ the energy minima look flat: $\lambda_{m}=0$. 
At low temperature, the expected equilibrium value of the energy is smaller than $E_t$.
Therefore, starting from high temperature conditions, the system does not 
manage to reach the equilibrium because the phase space becomes flatter and flatter in the vicinity of $E_T$. 
In the thermodynamic limit, 
the energy slowly drifts toward $E_T$ but never reaches it, which leads to aging. 
The mode-coupling
temperature corresponds to an equilibrium energy equal to $E_T$.
In a  finite size system, the
energy can cross the threshold so that the dynamics becomes activated. In this situation, the phenomenological 
trap models \cite{bouch,mezard}
give a fair description of the aging behavior. In particular, the regime just before
the thermalization is expected to be well described  \cite{junkurch} by the activated trap model \cite{bouch}.

The present Letter proposes a phenomenological modeling of the mode-coupling glass transition in the spirit
of the spherical $p$-sin model. On one hand, we aim at providing new statistical tools to describe, at least
at a mean-field level, the topology of an energy landscape, and more specifically the presence of an energy threshold.
On the other hand, the present model 
has a natural extension to finite
dimension, that we discuss in the conclusion. In this sense, we aim at giving a general framework to investigate the passage
from a mean-field description to a necessary real space description \cite{edi,garachan,berth} 
of both the glass transition and glassy dynamics.

%In finite dimension systems, $T_M$, a temperature issued from mean-field schemes, 
%becomes a crossover around which the global dynamics is not singular. 
%Recent studies have then focused on the spatial 
%structure of the dynamics and shed new light on
%the heteregeneous properties of the dynamics closed to the glass transition \cite{edi,ritort,garachan,berth}.

\begin{figure}
\includegraphics[width=3.8cm,height=2.4cm]{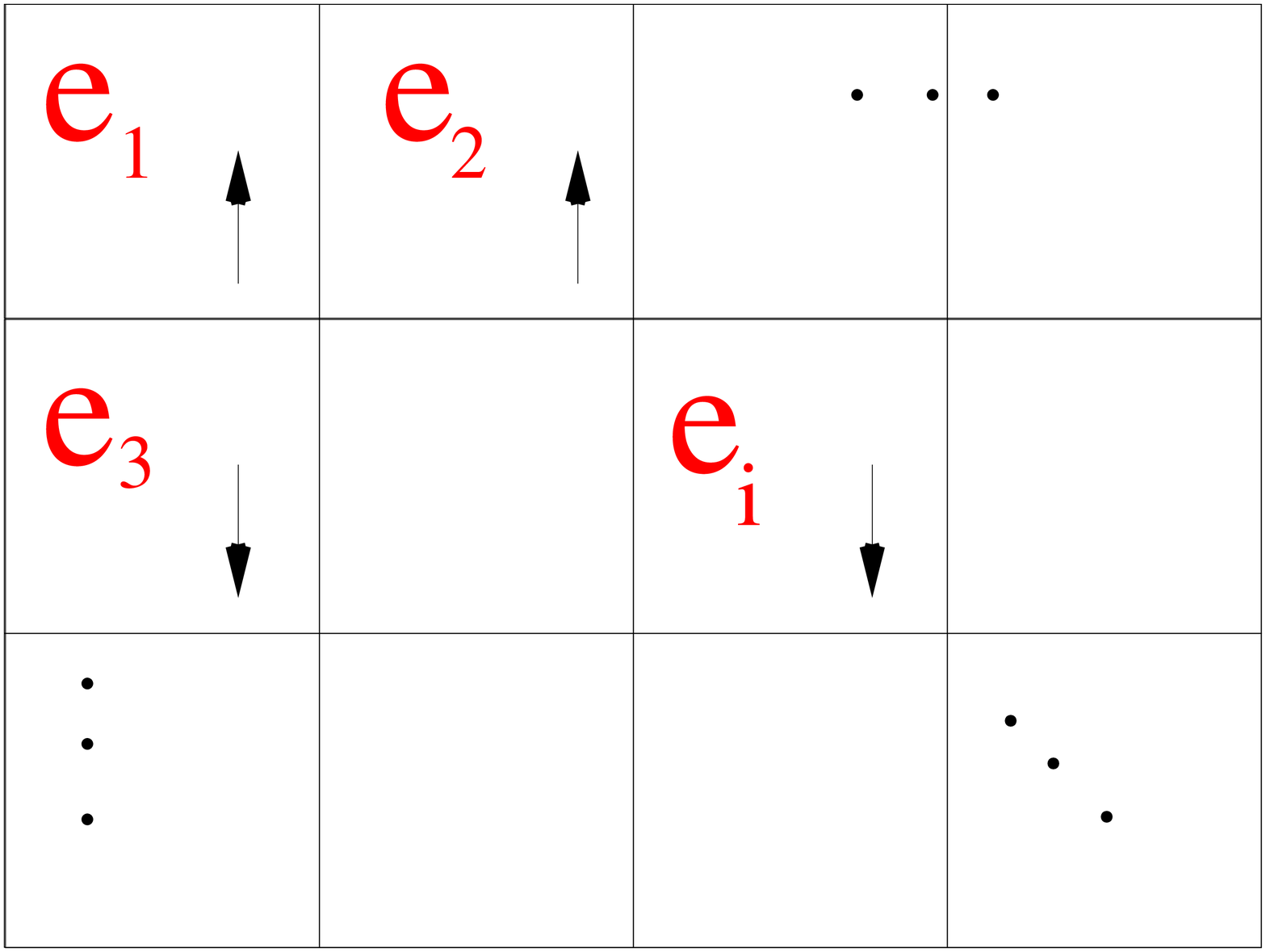}
\caption{Schematic view of the local energies. A random Ising-like spin ($\uparrow$, $\downarrow$  ) 
is initially assigned to each site $i$.}
\label{ei}
\end{figure}
%Indeed, it is not clear how to adapt MCT equations
%to finite dimension since there is no well-defined local field like the magnetization in ferromagnet's.
%On the other hand, the dynamics of trap models is not realistic since aging is due to energy barriers
%that scale like the size of the system. 
%Therefore, the effort to go to finite dimension has focused on systems 
%relatively different than the canonical mean-field models.

Unlike the trap models that deal with {\it global} energies,  our approach is based on the stochastic 
evolution of {\it local} energies
$\{e_i\}_{i=1...N}$. These energies
can be thought of as the energies of interaction felt by particles or spins, 
or as the energies of
mesoscopic regions just as in the facilitated spin approaches \cite{FA} --see Fig. \ref{ei}. For the sake of generality, we shall 
refer to the energy of a "site". To further measure the response of {\it neutral} observables \cite{Sollich}, i.e. 
uncorrelated with the energies, 
a random Ising-like spin $s_i$ 
is initially assigned to each site $i$. Next, $s_i$ flips {\it if and only if} 
a transition is accepted for $e_i$.

The $e_i's$ are supposed to evolve independently from each other. Thus, we drop the subindex $i$ and
deal with a {\it single-energy} $e$. In the spirit of the random energy model \cite{dede}, 
$e(t)$ takes values in an infinite ensemble $\mathcal{E}$ of independent random
energies distributed according to a Gaussian distribution $\rho(e)=\exp(-e^2)$ --where the energy unit is set to one.
Starting from an energy $e$ at time $t$, the dynamics consists in drawing
%unlike a random
%choice of the new energy $e'$ in $\mathcal{E}$ with probability $\rho(e')$, 
a new energy $e'$ following a 
connectivity function
$f_{\epsilon}(e\to e')$, and then, accepting it or not according to the Metropolis rule. The function $f_{\epsilon}$ 
depends on the
mean energy per site $\epsilon=\sum_i e_i/N$ and reflects the topological properties
of the energy surface at the global energy $N\epsilon$. 
Roughly speaking, $\epsilon$ can thus be thought of as an analog of the {\it self-consistent} magnetization in the 
ferro-magnetism {\it \`a la} Curie-Weiss.

\begin{figure}
\includegraphics[scale=0.25]{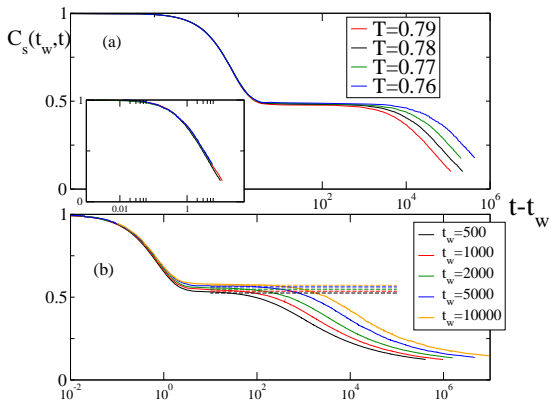}
\caption{Spin autocorrelation, $e_d=-0.7 \Rightarrow T_d \approx 0.71$. 
(a) Equilibrium curves. $N=2 \times 10^5$. Inset: the
time is rescaled by $(1-T/T_d)^{-4}$; the curves are normalized at the plateau value. 
(b) aging behavior. $T=0.6$. $N=5000$. Dashed lines are the plateau values predicted for
a full decoupling of the timescales (see text).
Similar curves are obtained for the energies. The times are measured in number of sweeps.}
\label{figAC}
\end{figure}

We shall endow $f_\epsilon$ with a specific dependence on $\epsilon$ to model the effect
of an energy threshold. In this case,
we find at low temperature a spin/energy relaxation with two time-sectors
(see Fig. \ref{figAC}). The aging scaling is 
fully determined by the evolution of $\epsilon$ and
the spin observable gives linear fluctuation dissipation relations (FDR) that lead to effective temperatures
larger than $T$. The energies have non-linear and non-monotonic FDR's.

\paragraph{\bf {Mimicking a dynamic glass transition}}

\begin{figure}
\includegraphics[width=6.5cm, height=3cm]{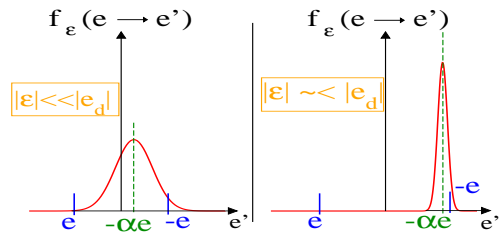}
\caption{
$\epsilon$-dependence of the connectivity function. For $\epsilon \sim e_d$, the evolution is essentially trapped within
the TSP's $\{-e,e\}$. The greater the exponent $\nu$, the more narrow $f_{\epsilon}$, i.e.
the more efficient the trapping within the TSP's. $e<0$ in this figure.}
\label{feps}
\end{figure}

$f_{\epsilon}$ is constrained by the following topological relation:
%so that glassy
%dynamics is not due to an explicit violation of this latter:
\ben
\rho(e)f_{\epsilon}(e \to e')&=&\rho(e')f_{\epsilon}(e' \to e)
\label{Con}
\een
This is the reciprocity property of the connectivity. It
says that the connection from a state $A$ to a state $B$ is identical to the connection
from $B$ to $A$. In the following,
we consider a Gaussian/$\delta$ function (see Fig. \ref{feps}):
\ben
f_{\epsilon} (e \to e')&=&((1-\epsilon /e_d)^{\nu}\pi )^{-\frac{1}{2}}
e^{-\frac{(e'+\alpha e)^2}{(1-\epsilon/e_d)^\nu}}
\; \mbox{if} \; \epsilon > e_d \quad
\label{f1}
\\
f_{\epsilon}(e\to e')&=&\delta(e'+e) \quad \mbox{if} \quad \epsilon \leq e_d
\label{f2}
\een
which verifies (\ref{Con}) if $\alpha=\sqrt{1-(1-\epsilon/e_d)^\nu}$.
$e_d$, an energy that does not depend
on the temperature and $\nu$, a positive exponent, are free parameters.

To justify this choice, let us consider the Monte-Carlo evolution of some disordered
spin model $\mathcal{S}$ where the spins are Ising-like. At each step, a spin is chosen with probability $1/N$ and flipped or not according
to the Metropolis rule. At infinite temperature, $\epsilon=0$ in our case, 
any transition is accepted. Then, between two successive
flips of the same spin, the global energy, and hence the local energy, strongly fluctuate.
We then expect a random choice for the local energies, which corresponds to
$f_{\epsilon=0}(e \to e') \sim
\exp(-e'^2)$ in our formalism. Close to a "mode-coupling" energy threshold, the scenario is different.
For instance, in the $p$-spin model the closer to the
threshold, the longer the system stays in the same region of the phase space \cite{laloux}. In other words,
fast fluctuations do not drift the system away from an initial configuration. In the spin model $\mathcal{S}$, 
fast fluctuations correspond to spin-flips and a configuration can be labeled by the data of its local energies. 
Therefore, close to the energy threshold, between two flips of the same spin, the corresponding local energy, $e$, should not
change much. A way to include this effect is to consider a trapping into a two-state process (TSP) with energies
$\{e,-e\}$. This justifies that $f_\epsilon(e \to e')$ becomes more and more peaked around $-e$ as 
$\epsilon \to e_d$, $e_d$ thus playing the role of the threshold --see Eq. (\ref{f1}) and Fig. \ref{feps}.
At the energy threshold, the system consists of an ensemble of TSP's that are described by the
$\delta$-function (\ref{f2}).

At high temperature, the Boltzmann equilibrium is recovered and  
$\epsilon=-1/2T$. Thus,
after a temperature quench below $T_d=1/2|e_d|$,
the energy first rapidly decreases toward $e_d$ and then, is slowed down by the trapping
of the TSP's. At low
temperature, 
we numerically find that the time needed 
to reach $e_d$ (measured in number of sweeps) is larger than $N$ when $\nu \geq 2$,
so that it diverges in the thermodynamic limit. 
%As announced,
%we have a threshold energy value, namely $e_d$, in the spirit of the $p$-spin model. 

\paragraph{\bf{Correlation functions}}

We have numerically investigated the connected energy/spin autocorrelation functions,
$C_e(t_w,t)=\sum_i (\langle e_i(t_w) e_i(t)\rangle-\langle e_i(t_w)\rangle \langle e_i(t) \rangle)/N$,
 and 
$C_s(t_w,t)=\sum_i (\langle s_i(t_w) s_i(t)\rangle-\langle s_i(t_w)\rangle \langle s_i(t) \rangle)/N$ respectively. The brackets
stand for an average over the noise history. 
The initial conditions are always taken at infinite temperature and 
the waiting time before measurements is noted $t_w$. 
 The simulations reported here were done with $\nu=4$ although similar 
results are obtained for different $\nu \geq 2$. 

%Fig. (\ref{figAC}) are the results of Monte-Carlo simulations where the spins follow the dynamics described above with
%Metropolis-like acceptance rates.
%The function $\alpha$ (a priori arbitrary chosen) is given by $1-\alpha(\epsilon)=(1-\epsilon/e_d)^4$ with $ed=-0.7$. 
%The reported quantity is
%the two-time autocorrelation function:
%
%\be
%C(t_w,t)=\langle \frac{1}{N} \sum_{i=1}^N s_i(t_w) s_i(t) \rangle
%\label{AC}
%\ee
%where by convention $t>t_w$.
%Note that spins and energies are uncorrelated since the evolution is exclusively based  
%on the dynamics of the stochastic energies $e_i$. Nevertheless,
%the autocorrelation (\ref{AC}) is the footprint of the 
%information lost between times $t_w$ and $t$. When $C=0$, the system has forgotten 
%the initial values of the $e_i$'s whereas $C>0$ means that the system still is influenced.
 
Fig. \ref{figAC} shows the spin autocorrelation at different temperatures. 
%(the curves for the energy 
%have exactly the same shape).
Three 
regimes must be distinguished. At high $T$ (data not shown), the system
exponentially relaxes and the
evolution is time translational invariant (TTI). At lower $T \gtrsim T_d$ (Fig. \ref{figAC}a), the autotocorrelation develops a 
plateau that becomes
TTI at finite time (not diverging with $N$). At further lower $T \leq T_d$, the system enters in 
the so-called aging regime (Fig. \ref{figAC}b). Then, we observe a short-time regime ($t-t_w \ll t_w$) that becomes TTI 
as $t_w$ increases  
whereas for long times $t-t_w \gg t_w$, the relaxation depends on $t_w$.

%A first hint to understand those behaviors is to compute the evolution of the total energy.
%At high temperature, it gaussianly fluctuates around its mean value $\epsilon=-N/2T$ (see Eq. (\ref{GB})+ 
%(\ref{ro})). At low temperature ($T<T_d=-1/2e_d$), it goes below the {\it threshold} value 
%$E_d =N e_d$ in a time that increases with $N$. Hence, in the thermodynamic limit, the system never penetrates the
%sub-threshold values $E \leq E_d$ ($\alpha \geq 1$) for any finite time and $\alpha$ tends to $1$ in the limit of 
%infinite time.

%As announced, our system has the peculiar features of the SpM: 
%i) at intermediate temperature, emergence of a plateau in the two-time autocorrelation function, ii) divergence of the 
%plateau length at a temperature $T_d$ (the so-called dynamical transition), iii) an energy threshold level below which the
%system never penetrates, 
%{\it the thermodynamic limit being taken before the long-time limit}, iv) a singular non-equilibrium behavior when $T<T_d$ 
%for which the
%two-time autocorrelation can be decomposed into two time-sectors, one that is TTI ($t-tw \sim \tau_0$) 
%and the longer one that ages, $t_w$ being then the relevant relaxation time.

These behaviors can be rationalized within a master equation (ME) approach. 
As $\epsilon \to e_d$, the distribution $f_\epsilon$ can be expanded in powers of $(1-\epsilon/e_d)^{\nu}$. Keeping only
the first terms, the ME for the energy $e$ reads \cite{jun}:
\ben
\nonumber
\partial_t P (e,t)&=&-w(e \to -e) P (e,t)+w(-e \to e) P (-e,t)
\\
 && \hspace*{-0.5cm} -(1-\epsilon(t)/e_d)^{\nu} F_1(P (\pm e),P '(-e),P ''(-e)) 
\label{dl}
\een
with $\epsilon(t)=\int de e P(e,t)$. $F_1(\cdot)$ is a functional of $P (x,t)$ and $w$ is
the Metropolis rate. 

When dealing with the above two-time autocorrelation functions, 
the initial conditions in Eq. (\ref{dl}) must be taken at time $t_w$. Thus,
the two first terms of the right hand side, that account for the relaxation within the TSP's,
contribute to the fast relaxation toward the plateau. 
The last term allows the
system to further relax with a typical timescale $\sim (1-\epsilon(t)/e_d)^{-\nu}$ that can be time-dependent if
$\epsilon(t)$ does not reach a stationary regime.
Therefore, at equilibrium the relaxation time for the decay from the plateau should scale like 
$(1-\epsilon/e_d)^{-\nu} \sim (T-T_d)^{-\nu}$ as $T \to T_d$ since the closer to $e_d$, the sharper the timescale separation
in (\ref{dl}).
Fig. \ref{figAC}a confirms this scaling law once normalized the plateau
values.

For $T \leq T_d$, $\epsilon(t)$ drifts
toward
$e_d$ without reaching it. 
In the limit $t_w \to \infty$ (rigorously taken after $N \to \infty$), fast and slow timescales totally
decouple. First, the TSP's locally equilibrate, which leads to
$P (e,t)w(e \to -e) = P (-e,t)w(-e \to e)$. Next, under these conditions the ME (\ref{dl}) reduces to:
\be
\partial_t P (e,t)=-(1-\epsilon(t)/e_d)^{\nu} F_2(P (e),P '(e),P ''(e)) 
\label{dl2}
\ee
where $F_2$ is a functional different than $F_1$ \cite{jun}. 
This gives the dynamical evolution of $P(e,t)$ during the decay from the plateau. The timescale  $(1-\epsilon(t)/e_d)^{-\nu}$
is now time-dependent. Furthermore, considering the decoupling of fast and slow timescales, the spin autocorrelation function in the aging
regime can be written as \cite{jun}:
\ben
C_s(t_w,t)&=&\int de_w de G(e,t|e_w,t_w)(p(e)-p(-e))
\label{Coag}
\\
\nonumber
&\times& (p(e_w)-p(-e_w))G(e_w,t_w)
\een
In this relation,
$G(e,t)$ is the energy density of the TSP's and is given by $(P(e,t)+P(-e,t))/2$. $G(e,t|e_w,t_w)$ is the
corresponding propagator with the initial condition 
$P(\pm e,t)=\delta(e\mp e_w)\delta(t-t_w)$. 
$p(e)$ 
is the Boltzmann weight restricted to the TSP $\{e,-e\}$. 
Thus, the relation (\ref{Coag}) means that the 
decay from the plateau comes 
from the decorrelation of the TSP's $\{e,-e\}$ due to a diffusion of the energy $e$ governed by the equation (\ref{dl2}).

\begin{figure}
\includegraphics[scale=0.2]{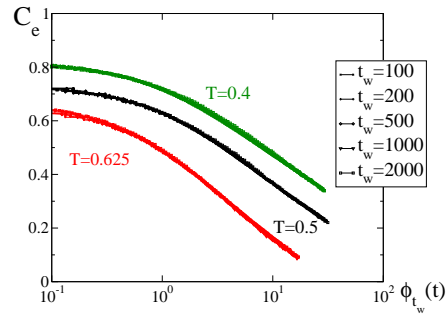}
\caption{Energy autocorrelation {\it vs.} $\phi_{t_w}(t)$. $N=4000$. 
$e_d=-0.8 \Rightarrow  T_d=0.625$. For each simulation, $t$ runs from $t_w$ to $1000 \times t_w$. 
The curves are normalized at the plateau value with the shortest $t_w$. 
$(1-\epsilon(t)/e_d)^{-4}$ was fitted by stretched exponentials 
to obtain $\phi_{t_w}(t)$. Exponents of the stretched exponentials are pretty similar and are around $0.1$.
The fit correlation coefficient was always above 0.9995. Similar results are obtained for the spin autocorrelation.} 
\label{AC2}
\end{figure}

Denoting $\phi_{t_w}(x)=\int_{t_w}^{x} dt'(1-\epsilon(t')/e_d)^{-\nu}$, Eq. (\ref{dl2}) calls for a solution 
$G(e,t|e_w,t_w)=\tilde G(e,\phi_{t_w}(t)|e_w,0)
$. Inserting this relation into (\ref{Coag})
and noticing that $G(e,t_w)$ becomes stationary as $t_w \to \infty$, the aging scaling reads:
\be
C_{s,e}(t_w,t)=\tilde C_{s,e}(\phi_{t_w}(t))
\ee
the reasoning for the energies being identical. This scaling is numerically well verified (see Fig. \ref{AC2}), 
which corroborates our treatment of the aging regime. Moreover, in this scope, the
plateau value for the spins at time $t_w$ is equal to $\int de \; G(e,t_w)(p(e)-p(-e))^2$, that
is also numerically well verified (see Fig. \ref{figAC}).
The Edwards-Anderson parameter $q_{EA}$ is given by this quantity when $t_w \to \infty$.

To summarize, two distinct timescales appear at low temperature. A fast one coming from the relaxation within the TSP's,
and a slower one that increases as $\epsilon(t)$ drifts to $e_d$. Notice that
a full aging regime ($t/t_w$ scaling) is expected only if $(1-\epsilon(t)/e_d)^{-\nu}$ decreases as $1/t$ since then
$\phi_{t_w} \propto \log(t/t_w)$.
However, we have never seen this regime in our simulations.

\paragraph{\bf Response to a field}
\begin{figure}
\includegraphics[scale=0.27]{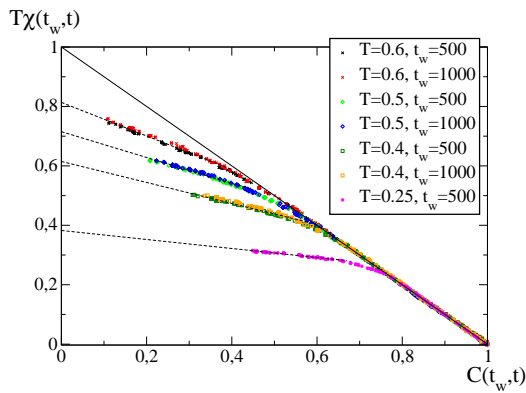}
\caption{FDR. $N=1000$, $e_d=-0.7$. The solid line is the equilibrium relation. The responses have been computed using $h=0.1$.}
\label{figFDR}
\end{figure}
The two-time magnetic susceptibility corresponding to the spin autocorrelation function
reads
%while dissipation results in a delay of the system to
%respond to external perturbation. It is described by the two-time
%susceptibility 
$\chi_s(t_w,t)=\partial \langle \xi s(t) \rangle/\partial h(t_w)|_{h=0}$. 
$\langle \xi s(t) \rangle=\frac{1}{N}\sum \langle \xi_i s_i(t) \rangle$ is measured a time $t-t_w$ after having switched on,
at time $t_w$, a small constant field $h$ coupled to $\xi s$. 
The $\xi_i$'s are quenched random variables that take values $\pm 1$  \cite{parisi}.
%In most of the aging systems, $\chi(t_w,t)$ ages with the same timescale as the 
%correlation function. 
At low temperature and in the long-time limit, 
a piecewise relation between $\chi_s$ and $C_s$ has been found in the $p$-spin model \cite{cugkurch1}:
%Remarkably, in both theoretical, experimental and numerical situations, it has been observed that 
%the $t$-parametric plot $(C(t_w,t),\chi(t_w,t))$ leads to a two-temperature scenario at large $t_w$:
\ben
\label{FDR1}
\hspace*{-0.5cm} \chi_s(t_w,t) &=& (1-C_s(t_w,t))/T \; \; {\mbox if} \; \; C_s>q_{EA} \\
\label{FDR2}
&=&(q_{EA}-C_s(t_w,t))/T_e \; \; {\mbox if} \; \; C_s<q_{EA}
\een
The relation (\ref{FDR1}) is the usual equilibrium relation and is due to a local equilibrium property of the system. 
The linear relation (\ref{FDR2}) in the aging regime
leads to an effective temperature $T_{e} \geq T$ that shares the thermodynamic properties of canonical temperatures 
\cite{kur,cugkurch2}. 
Interestingly,
such a behavior has been later observed in numerical simulations and in experiments as well \cite{criri}, which prevents
any artifact of the mean-field treatment.

We computed $\chi_s$ in our system by adding to $e_i$ a magnetic energy $e_m=-h \xi_i s_i$.
Fig. \ref{figFDR} shows,  for different $t_w$, the plot $T\chi(t_w,t)=u(C(t_w,t))$ obtained in the linear 
regime of the response. We see that it can be divided 
into two sectors corresponding to the fast relaxation and the aging regime respectively. 
The former gives an equilibrium-like relation. This directly results from the local 
equilibrium within the TSP's (see above). The interesting results come 
from the
aging part since it seems that we have a {\it linear} FDR for which we can define an effective temperature
$T_e \geq  T$. As in the $p$-spin models, $T_e$ slightly increases as $T$ decreases. 

We have also investigated the response of the local energies. We found non-monotonic behaviors in the aging regime \cite{jun} 
similarly to the results of \cite{mayer} in finite dimension kinetic constrained models. Thus, our results suggest
that even in mean-field situations, local energies may respond non-monotonically which may lead in some case to negative $T_e$'s 
\cite{mayer}.

%
%\section{Conclusion}
%

\paragraph{\bf Conclusion} We have considered the stochastic dynamics of $N$ independent {\it local} energies. 
At each Monte-Carlo step, the new energies are drawn from a distribution $f_{\epsilon}$ whose properties {\it self-consistently}
depend on the {\it global} energy $N\epsilon$ of the system. The dependence is chosen in order to model
an energy threshold in the spirit of the $p$-spin model.
Given this mode-coupling scenario, 
we obtain at low temperature the typical properties of aging. 
In
particular, our approach gives a
rather intuitive (and simple) illustration of how two time-sectors may appear in a mean-field situation. 
The linear violation of the fluctuation dissipation relation, for observables uncorrelated with the energies, further 
confirms the similarity between our {\it phenomenological} description and the {\it microscopic} disordered models.

Interestingly, one can generalize this mean-field approach to finite dimension replacing $\epsilon$ by a local energy 
$\epsilon_i=\frac{1}{z}\sum_{<j>} e_j$
where the sum is taken over the $z$ nearest neighbors. In this case, one expects $T_d$ to become a crossover below which
dynamical heterogeneities should play an important role. 
%Interestingly, the differences between finite dimension and mean-field
%approach might be studied in a unified way \cite{jun2} within this scope.    

I am pleased to thank R. Agra, J. Kurchan, F. Ritort, P. Sollich and S. Tanase-Nicola
for illuminating discussions. I acknowledge financial support
from the European network STIPCO, Grant No. HPRNCT200200319.

\end{document}